\documentclass[aps,pre,twocolumn,groupedaddress,showpacs]{revtex4}

\usepackage{amsmath,graphicx}

\begin{document}

\title{Analytical results on the Muller's ratchet effect in growing populations}

\author{Leonardo P. Maia}
\email[]{lpmaia@ifsc.usp.br}

\affiliation{Instituto de F\'{i}sica de S\~ao Carlos, Universidade de S\~ao Paulo, Caixa Postal 369, 13560-970 S\~ao Carlos SP, Brazil}

\date{\today}

\begin{abstract}
Fontanari {\it et al} introduced [{\it Phys. Rev. Lett.} {\bf 91}, 218101 (2003)] a model for studying the Muller's ratchet phenomenon in growing asexual populations. They studied two situations, either including or not a death probability for each newborn, but were able to find analytical (recursive) expressions only in the no-decay case. In this paper a branching process formalism is used to find recorrence equations that generalize the analytical results of the original paper besides confirming the interesting effects their simulations revealed.
\end{abstract}


\maketitle

It is widely recognized that the rate of deleterious mutations being much higher than that of either reverse or beneficial mutations can be a serious threat to the survival of populations at the molecular level. About forty years ago, H. J. Muller conjectured \cite{muller-1964} that, in such conditions, the mean fitness of finite lineages lacking mechanisms of genetic repair should decay with time, due to the sucessive loss of the fittest individuals. This stepwise fluctuation-induced phenomenon is known as Muller's ratchet \cite{felsenstein-1974} and have received growing attention in recent years. It has been argued to be responsible for the origin of some diseases, for the fitness decay in some experiments with microorganisms and even for the origin of sex, as summarized in \cite{baake-gabriel-2000, drossel-2001}, although many of these claims still lack conclusive evidence.

In contrast with the first studies \cite{haigh-1978}, recent work on Muller's ratchet have focused on models with variable population size, either under growing conditions \cite{manrubia-et-al-2003b, fontanari-et-al-2003} or mixing growth with bottlenecks \cite{colato-fontanari-2001, gordo-dionisio-2005}. Indeed, this is the realistic condition that holds both in nature and in controlled experiments \cite{elena-lenski-2003}, with exponential growth of microorganisms and bottleneck transfers.

Specifically, in \cite{fontanari-et-al-2003} Fontanari and collaborators (henceforth FCH) introduced a fully stochastic discrete time model for growing asexual lineages in which each member of the population is replaced in the next generation by a number $k$ of descendants distributed according to a Poisson distribution of parameter $\lambda=R$, i.e., with probability
\begin{eqnarray}
p(k ; \lambda) = \exp{(-\lambda)} \frac{\lambda^k}{k!}.
\end{eqnarray}
Each newborn acquires a number of new mutations also Poisson distributed, but now with mean value $U$. In addition, an individual with $j$ mutations has a chance to leave offspring only if it survives, what happens with probability $(1-s)^j$, where $s \in [0,1]$ is a selective coefficient. FCH focused on finding conditions for the halting of Muller's ratchet when the population is founded by a single mutation-free individual (called master sequence). They used three recursion equations to study the neutral case $s=0$ and recurred to simulations in the case of evolution under decay.

The aim of this Brief Report is to show that the theory of branching stochastic processes, both in its simple (just one type of individual) and multi-type versions, allows a complete description of FCH's model, not only recovering their analytical results in the neutral case in a straightforward way, but also giving expressions valid for the case $s>0$. In particular, it is shown that the counterintuitive acceleration of the ratchet's activity with an increase on the selective coefficient found through simulations in \cite{fontanari-et-al-2003} is a real phenomenon and not an artifact. The theory of branching processes is a veritable subject, originally developed by J. B. S. Haldane \cite{haldane-1927} and R. A. Fisher \cite{fisher-1922,fisher-1930} in the birth of the modern population genetics and today is described in many textbooks like \cite{karlin-taylor-1975}. The multi-type theory have found lots of applications in evolutionary dynamics, \textit{e.g.} \cite{barton-1995}.

To start with, we consider the neutral case. Defining $\mathcal{N} (k,t)$ as the probability that the population is composed of $k$ individuals in generation $t$, regardless their mutational load, FCH found that the generating function $g(z,t) = \sum_{k=0}^{\infty} z^k \mathcal{N} (k,t)$ obeys the equation
\begin{eqnarray}
g(z,t+1) = g[e^{-R(1-z)},t].
\label{eq:gdeles}
\end{eqnarray}
Being $\mathcal{S}^{(t)}$ the probability that the population is not extinct at generation $t$, $\mathcal{S}^{(t)} = 1-g(0,t)$ and they argued that
\begin{eqnarray}
\mathcal{S}^{(t+1)} = 1-\exp{(-R \mathcal{S}^{(t)})}.
\label{eq:Sdeles}
\end{eqnarray}
In a general simple branching process (henceforth SBP), if there is a single founder and $\varphi(z)$ is the generating function of the progeny distribution $f(k)$, $\varphi(z) = \sum_{k=0}^{\infty} z^N f(k)$, then
\begin{eqnarray}
g(z,t+1) = g[\varphi(z),t] = \varphi[g(z,t)]
\label{eq:recur-sbp}
\end{eqnarray}
and it is easily seen that Eqs. (\ref{eq:gdeles}) and (\ref{eq:Sdeles}) are just particular instances of the above equation in the Poissonian case, $f(k) = p(k ; R)$.

It is another standard result in the theory of the SBP that the mean number of offspring of an individual been strictly greater than 1 is a necessary and sufficient condition for the asymptotic survival probability of the population to be greater than zero. In the neutral case, neglecting the genotype of all individuals is harmless: they can be considered all equivalent. Thus, the evolution of the system is properly described by a SBP and, when $s=0$, a necessary condition for the survival of the population is $R>1$, a result FCH derived from Eq. (\ref{eq:Sdeles}).

More results follow from a decomposition theorem \cite{cinlar-1975} that is essential to this work. Roughly speaking, it states that if each ``object'' generated by a Poisson process with mean $R$ is independently attributed to a class with probability $p_i$, then the number of objects in class $i$ is again Poisson distributed, with parameter $R p_i$, and is independent of other classes. This ``allocation'' is exactly what mutation does and the classes are defined by the mutational load.

Being $e^{-U}$ the probability of a perfect replication, each individual originates an average of $R e^{-U}$ error-free replicas of itself, following a Poissonian SBP. Therefore, if $R e^{-U} \le 1$, the mutation-free subpopulation certainly goes extinct. Remembering we still are in the no-decay case, this argument applies qualitatively to all classes of individuals, even whether a new class that had just became the least-loaded one has more than one member at that moment. Hence, if $R>1$ but $R e^{-U} \le 1$, all classes of sequences have a finite lifetime and Muller's ratchet never halts in the surviving populations. Conversely, if $R e^{-U} > 1$, there is a positive probability $\mathcal{S}^{(\infty)}$ that the least-loaded class survives the expansion process, given by the asymptotic solution of Eq. (\ref{eq:Sdeles}) when $R$ is replaced by $R e^{-U}$. Every time the ratchet clicks is equivalent to a new start of the process, always with a positive survival probability. Hence, if the population does not get extinct, some class will survive indefinitely and halt the ratchet.

FCH noticed the criterium they found for the halting of Muller's ratchet ($U < \ln R$ or, equivalently, $R e^{-U} > 1$) implies that, on average, each master sequence must generate more than one perfect replica in order to the population to be viable. However, despite recognizing that criterium as intuitive, they could not predict it in advance. Actually, they derived it from a recursion for $\mathcal{P}_n(t)$, the probability that at generation $t$ the minimum number of mutations in the population be $n$, valid only in the neutral case. It turns out to be the most relevant quantity for the study of Muller's ratchet since it reveals the time dependence of the mean number of mutations of the fittest member of the population. So it is noteworthy that the theory of a multi-type branching process allowed the finding of the main result of this paper, an analogous recursion for $\mathcal{P}_n(t)$ in the general case $s \in [0,1]$, which will be discussed from now on. To be honest, there is an infinite number of types of individuals, since there is no such thing as a maximal mutational load. But the results inspire confidence in the adopted approach.

At this point, some remarks on notation are necessary. A vector $\mathbf{u}$ has infinite components and the first index is $0$, to account for the mutation-free class. Explicitly, a phrase like ``the population is in state $\mathbf{k}$'' means that $\mathbf{k}=(k_0,k_1,\ldots)$ and there are $k_j$ individuals in class $j$. It is also convenient to define $\mathbf{u} = (\mathbf{u}^{(j)}, \overline{\mathbf{u}^{(j)}})$, where $\mathbf{u}^{(j)}=(u_0,\ldots,u_{j-1})$ and $\overline{\mathbf{u}^{(j)}}=(u_j,u_{j+1},\ldots)$ for any $j \ge 1$. The same rules apply to constant vectors too. So, $\mathbf{0}^{(j)}$ means a vector with just $j$ components, all null, and, despite how close $\overline{\mathbf{1}^{(j)}}$ appears to be to $\mathbf{1}$, they are not the same object, since the first index referred to in $\overline{\mathbf{1}^{(j)}}$ is $j$.

Let $f_i(\mathbf{k})$ be the probability of an $i$-individual to generate offspring $\mathbf{k}$ and $\mathcal{N}_i (\mathbf{k},t)$ be the probability that the population is in state $\mathbf{k}$ at generation $t$, given a single founder with $i$ mutations. The generating functions associated to these two joint distributions are
\begin{eqnarray}
\varphi_i(\mathbf{z}) = \sum_{\mathbf{k}} f_i (\mathbf{k}) \prod_{j=0}^{\infty} z_j^{k_j},
\end{eqnarray}
and
\begin{eqnarray}
g_i(\mathbf{z},t) = \sum_{\mathbf{k}} \mathcal{N}_i (\mathbf{k},t) \prod_{j=0}^{\infty} z_j^{k_j},
\label{eq:gi}
\end{eqnarray}
respectively. The generalization of Eq. (\ref{eq:recur-sbp}) to the multi-type setting is
\begin{eqnarray}
\mathbf{g} (\mathbf{z},t) = \boldsymbol{\varphi} [\mathbf{g}(\mathbf{z},t-1)],
\label{eq:recuritbp}
\end{eqnarray}
when there is only one founder. Thus, to characterize the population at a given time, whatever be the initial condition, it is necessary to know its properties in the preceding generation as if it had originated from all possible types of founders.

Inquiries about extinction acquire a broder sense in this case, since now it is possible to talk about the survival of specific classes inside the community. Let $q_i(j,t)$ be the probability that the smallest index of a populated class at time $t$ be at least $j$, given a founder in class $i$. From Eqs. (\ref{eq:gi}) and (\ref{eq:recuritbp}),
\begin{eqnarray}
\mathbf{q}(j,t) = \mathbf{g}[(\mathbf{0}^{(j)}, \overline{\mathbf{1}^{(j)}}),t] = \boldsymbol{\varphi}[\mathbf{q}(j,t-1)].
\label{eq:qjtbf}
\end{eqnarray}
It is clear that $\mathcal{P}_{ij}(t)$, the probability distribution of the smallest index $j$ of a class still alive at $t$, given a founder in class $i$, is given by
\begin{eqnarray}
\mathcal{P}_{ij}(t) = q_i(j,t) - q_i(j+1,t),
\label{eq:diferenca}
\end{eqnarray}
and, since $\mathcal{P}_n (t) = \mathcal{P}_{0n}(t)$, this is all the information needed to study Muller's ratchet.

The existence of analytical expressions for the multidimensional generating functions is essential for this proposal to be useful. At this point, the decomposition theorem of Poisson processes enters again. It is important to notice that, under decay, each mutant is temporarily allocated on some class, depending on its mutational load, but after that it may not survive and thus may be ``redirected'' to a ``sink class'' that plays no role on the dynamics. Therefore, the factor $p_i$ informally introduced in the presentation of the decomposition theorem must take into account both mutation and the survival probability, while the number of individuals in any class still is Poisson distributed with parameter $R p_i$. For instance, the number of individuals with $i$ mutations directly descending from a $j$-mutant is given by a Poisson distribution with mean $R \, p(i-j; U) (1-s)^i$. Of course, independence of classes still holds and this allows a convenient factorization of the joint distribution $f_i(\mathbf{k})$ as the product of Poisson distributions. Consequently, $\varphi_i(\mathbf{k})$ also factorizes (as a product of Poissonian generating functions) and, after some algebra, it follows from Eq. (\ref{eq:qjtbf}) that
\begin{eqnarray}
q_i(j,t)  = \prod_{k=i}^{j-1} \exp \left\{ R \, p(k-i; U) \, (1-s)^k \right. \cdot
\nonumber \\
\cdot \left. \left[ q_k(j,t-1)-1 \right] \right\}.
\label{eq:qijtmain}
\end{eqnarray}
The product is finite because $q_i(j,t) \equiv 1$ when $j \le i$. Some time after discovering this result, we were informed that it is a special case of a general theory developed in \cite{barton-1995} and \cite{johnson-barton-2002}. Nonetheless, its present application to the study of Muller's ratchet in FCH's model is still unreported.

\begin{figure}
\includegraphics{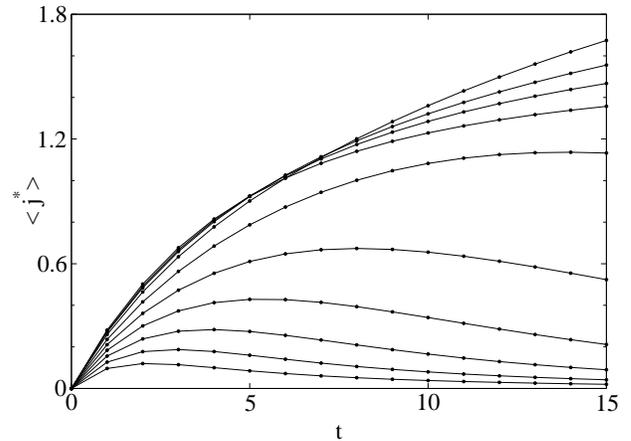}
\caption{\label{fig:acelera} Dynamical behavior of the average mutational load of the least loaded class for $R=2$ and $U=0.6$. At $t=15$, from top to bottom, $s=0.1$, $0.05$, $0.03$, $0.0$, $0.2$ and from this value $s$ increases to $0.7$ in steps of $0.1$. The anomalous activity of Muller's ratchet found by FCH appears clearly. The continuous lines just aid visualization of the discrete time dynamics.}
\end{figure}

\begin{figure}
\includegraphics{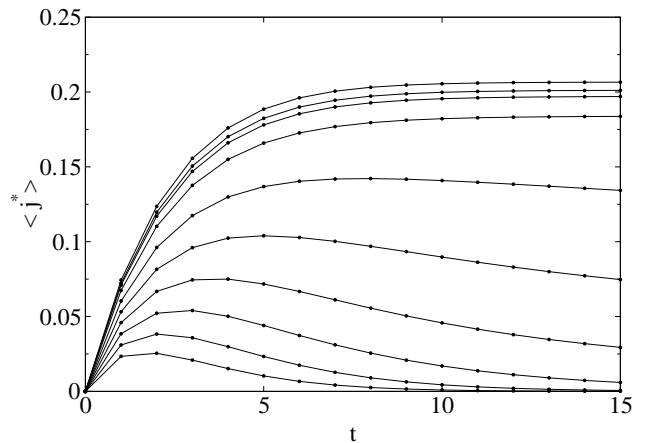}
\caption{\label{fig:nao-acelera} Dynamical behavior of the average mutational load of the least loaded class for $R=2$ and $U=0.2$. At $t=15$, from top to bottom, $s=0.0$, $0.03$, $0.05$, $0.1$ and from this value $s$ increases to $0.7$ in steps of $0.1$. Here, Muller's ratchet presents a monotonic dependence on $s$. The lines are as in Fig. \ref{fig:acelera}.}
\end{figure}

When $s=0$, $q_i(j,t)$ depends only on $j-i$ and Eqs. (\ref{eq:diferenca}) and (\ref{eq:qijtmain}) together are equivalent to the recursion for $\mathcal{P}_n (t)$ derived in \cite{fontanari-et-al-2003}, as expected. And, under decay, they still can be iterated easily. As an example, Fig. \ref{fig:acelera} (similar to Fig. 2 in \cite{fontanari-et-al-2003}) illustrates an unexpected phenomenon discovered by FCH: at a given generation, it is possible that the average mutation load of the least-loaded class increases if selection gets stronger, $R$ and $U$ kept fixed. Indeed, this effect seems counterintuitive at first. But since fitness is absolute in the model, any explanation must not rely on competition in the population and the analysis may not be straightforward. It seems valid for the author to reverse the reasoning: why should an increase in the intensity of decay always imply a decrease in the average mutational load of the fittest class in such a complex model, in which fitness is absolute and any change in parameters affect the probability of extinction? It is important to remember that all averages are calculated with probabilities conditionated on the survival of the population. Maybe the enhanced activity of Muller's ratchet seen in Fig. \ref{fig:acelera} be just a consequence of some peculiar dynamics of the spectrum of the population (regarding the distribution of individuals among the mutational classes) in some region of the parameter space. If $R$ is not big and mutation is high enough, the lower classes contribute mostly to higher ones, get less populated and, consequently, are more sensible to fluctuations. This scenery seems appropriate for an abrupt loss of lower classes and that is exactly what results from an empirical study of some combinations of parameters, that the ocurrence of the anomalous behavior in Muller's ratchet is favored by high mutation rates and low fertility. For instance, Fig. \ref{fig:nao-acelera}, where the mutation rate is just a bit lower than in Fig. \ref{fig:acelera}, already shows a monotonic response of Muller's ratchet to variation in $s$.

Finally, the SBP theory gives the condition for the halting of Muller's ratchet also under decay. The reasoning is analogous to the neutral case. By definition, if the ratchet does not halt, the minimum number of mutations in the system grows unrestrictly. In this case, since $(1-s)^j$ decreases monotonically to zero with the mutational load, no matter how small is $s>0$, there is a finite time when the minimal mutational load $j^*$ is so high that the average number of perfect replicates of each of the fittest individuals, $R e^{-U} (1-s)^{j^*}$, is smaller than 1 and even the fittest subpopulation certainly goes extinct, and obviously extinction is the fate of the whole population. Thus, no population can stand endless mutation accumulation, and the ratchet is certainly halted in any indefinitely surviving lineage. Besides that, $R e^{-U} > 1$ turns out to be the necessary condition for survival under decay, since the master class is unaffected by $s$ and performs better than all other classes in surviving.

Hence, whatever be $s$, the halting criterium of Muller's ratchet is $R e^{-U} > 1$, although it does not assure the survival of the population. It is noteworthy this result is a direct consequence of the basic theory of branching processes. But this formalism also gives elaborate quantitative tools that allow a thorough study of models of mutation accumulation in growing lineages. In particular, any decay law can be analyzed and the assumptions concerning the asymptotic survival of individuals with few mutations now can be evaluated at low computational cost.

Besides that, the stochastic dynamics of a growing lineage described in the present paper, as well as analogous dynamical solutions recently found for the deterministic behavior of infinite populations evolving on multiplicative \cite{maia-et-al-2003} and truncated \cite{maia-2005} fitness landscapes, may prove useful in the construction of general theoretical models (an early example is \cite{colato-fontanari-2001}) suited for describing populations mixing growth with bottlenecks, like that evolved under the serial transfer protocol of experimental evolution \cite{chao-1990, elena-lenski-2003}.

These important themes are out of the scope of this short paper and will be discussed elsewhere.

\begin{acknowledgments}
This research was supported by the Brazilian agency FAPESP and innitiated when the author worked on both his current address and at Instituto de Ci\^encias Matem\'aticas e de Computa\c{c}\~ao, Universidade de S\~ao Paulo, S\~ao Carlos-SP, Brazil.
\end{acknowledgments}


\begin{thebibliography}{12}
\expandafter\ifx\csname natexlab\endcsname\relax\def\natexlab#1{#1}\fi
\expandafter\ifx\csname bibnamefont\endcsname\relax
  \def\bibnamefont#1{#1}\fi
\expandafter\ifx\csname bibfnamefont\endcsname\relax
  \def\bibfnamefont#1{#1}\fi
\expandafter\ifx\csname citenamefont\endcsname\relax
  \def\citenamefont#1{#1}\fi
\expandafter\ifx\csname url\endcsname\relax
  \def\url#1{\texttt{#1}}\fi
\expandafter\ifx\csname urlprefix\endcsname\relax\def\urlprefix{URL }\fi
\providecommand{\bibinfo}[2]{#2}
\providecommand{\eprint}[2][]{\url{#2}}

\bibitem[{\citenamefont{Muller}(1964)}]{muller-1964}
\bibinfo{author}{\bibfnamefont{H.~J.} \bibnamefont{Muller}},
  \bibinfo{journal}{Mutat. Res.} \textbf{\bibinfo{volume}{1}},
  \bibinfo{pages}{2} (\bibinfo{year}{1964}).

\bibitem[{\citenamefont{Felsenstein}(1974)}]{felsenstein-1974}
\bibinfo{author}{\bibfnamefont{J.}~\bibnamefont{Felsenstein}},
  \bibinfo{journal}{Genetics} \textbf{\bibinfo{volume}{78}},
  \bibinfo{pages}{737} (\bibinfo{year}{1974}).

\bibitem[{\citenamefont{Baake and Gabriel}(1999)}]{baake-gabriel-2000}
\bibinfo{author}{\bibfnamefont{E.}~\bibnamefont{Baake}} \bibnamefont{and}
  \bibinfo{author}{\bibfnamefont{W.}~\bibnamefont{Gabriel}}, in
  \emph{\bibinfo{booktitle}{Ann. Rev. Comp. Phys.}}, edited by
  \bibinfo{editor}{\bibfnamefont{D.}~\bibnamefont{Stauffer}}
  (\bibinfo{publisher}{World Scientific}, \bibinfo{address}{Singapore},
  \bibinfo{year}{1999}), vol.~\bibinfo{volume}{9}, pp.
  \bibinfo{pages}{203--264}.

\bibitem[{\citenamefont{Drossel}(2001)}]{drossel-2001}
\bibinfo{author}{\bibfnamefont{B.}~\bibnamefont{Drossel}},
  \bibinfo{journal}{Adv. Phys.} \textbf{\bibinfo{volume}{50}},
  \bibinfo{pages}{209} (\bibinfo{year}{2001}).

\bibitem[{\citenamefont{Haigh}(1978)}]{haigh-1978}
\bibinfo{author}{\bibfnamefont{J.}~\bibnamefont{Haigh}},
  \bibinfo{journal}{Theor. Pop. Biol.} \textbf{\bibinfo{volume}{14}},
  \bibinfo{pages}{251} (\bibinfo{year}{1978}).

\bibitem[{\citenamefont{Manrubia et~al.}(2003)\citenamefont{Manrubia, L\'azaro,
  P\'erez-Mercader, Escarm\'{\i}s, and Domingo}}]{manrubia-et-al-2003b}
\bibinfo{author}{\bibfnamefont{S.~C.} \bibnamefont{Manrubia}},
  \bibinfo{author}{\bibfnamefont{E.}~\bibnamefont{L\'azaro}},
  \bibinfo{author}{\bibfnamefont{J.}~\bibnamefont{P\'erez-Mercader}},
  \bibinfo{author}{\bibfnamefont{C.}~\bibnamefont{Escarm\'{\i}s}},
  \bibnamefont{and} \bibinfo{author}{\bibfnamefont{E.}~\bibnamefont{Domingo}},
  \bibinfo{journal}{Phys. Rev. Lett.} \textbf{\bibinfo{volume}{90}},
  \bibinfo{pages}{188102} (\bibinfo{year}{2003}).

\bibitem[{\citenamefont{Fontanari et~al.}(2003)\citenamefont{Fontanari, Colato,
  and Howard}}]{fontanari-et-al-2003}
\bibinfo{author}{\bibfnamefont{J.~F.} \bibnamefont{Fontanari}},
  \bibinfo{author}{\bibfnamefont{A.}~\bibnamefont{Colato}}, \bibnamefont{and}
  \bibinfo{author}{\bibfnamefont{R.~S.} \bibnamefont{Howard}},
  \bibinfo{journal}{Phys. Rev. Lett.} \textbf{\bibinfo{volume}{91}},
  \bibinfo{pages}{218101} (\bibinfo{year}{2003}).

\bibitem[{\citenamefont{Colato and Fontanari}(2001)}]{colato-fontanari-2001}
\bibinfo{author}{\bibfnamefont{A.}~\bibnamefont{Colato}} \bibnamefont{and}
  \bibinfo{author}{\bibfnamefont{J.~F.} \bibnamefont{Fontanari}},
  \bibinfo{journal}{Phys. Rev. Lett.} \textbf{\bibinfo{volume}{87}},
  \bibinfo{pages}{238102} (\bibinfo{year}{2001}).

\bibitem[{\citenamefont{Gordo and Dionisio}(2005)}]{gordo-dionisio-2005}
\bibinfo{author}{\bibfnamefont{I.}~\bibnamefont{Gordo}} \bibnamefont{and}
  \bibinfo{author}{\bibfnamefont{F.}~\bibnamefont{Dionisio}},
  \bibinfo{journal}{Phys. Rev. E} \textbf{\bibinfo{volume}{71}},
  \bibinfo{pages}{031907} (\bibinfo{year}{2005}).

\bibitem[{\citenamefont{Elena and Lenski}(2003)}]{elena-lenski-2003}
\bibinfo{author}{\bibfnamefont{S.~F.} \bibnamefont{Elena}} \bibnamefont{and}
  \bibinfo{author}{\bibfnamefont{R.~E.} \bibnamefont{Lenski}},
  \bibinfo{journal}{Nature Rev. Genet.} \textbf{\bibinfo{volume}{4}},
  \bibinfo{pages}{457} (\bibinfo{year}{2003}).

\bibitem[{\citenamefont{Haldane}(1927)}]{haldane-1927}
\bibinfo{author}{\bibfnamefont{J.~B.~S.}~\bibnamefont{Haldane}},
  \bibinfo{journal}{Proc. Camb. Philos. Soc.} \textbf{\bibinfo{volume}{23}},
  \bibinfo{pages}{838} (\bibinfo{year}{1927}).

\bibitem[{\citenamefont{Fisher}(1922)}]{fisher-1922}
\bibinfo{author}{\bibfnamefont{R.~A.}~\bibnamefont{Fisher}},
  \bibinfo{journal}{Proc. R. Soc. Edinb.} \textbf{\bibinfo{volume}{42}},
  \bibinfo{pages}{321} (\bibinfo{year}{1922}).

\bibitem[{\citenamefont{Fisher}(1930)}]{fisher-1930}
\bibinfo{author}{\bibfnamefont{R.~A.}~\bibnamefont{Fisher}},
  \emph{\bibinfo{title}{The Genetical Theory of Natural Selection}}
  (\bibinfo{publisher}{Clarendon Press}, \bibinfo{address}{Oxford},
  \bibinfo{year}{1930}), \bibinfo{edition}{1st} ed.

\bibitem[{\citenamefont{Karlin and Taylor}(1975)}]{karlin-taylor-1975}
\bibinfo{author}{\bibfnamefont{S.}~\bibnamefont{Karlin}} \bibnamefont{and}
  \bibinfo{author}{\bibfnamefont{H.~M.} \bibnamefont{Taylor}},
  \emph{\bibinfo{title}{A First Course in Stochastic Processes}}
  (\bibinfo{publisher}{Academic Press}, \bibinfo{address}{New York},
  \bibinfo{year}{1975}), \bibinfo{edition}{1st} ed.

\bibitem[{\citenamefont{Barton}(1995)}]{barton-1995}
\bibinfo{author}{\bibfnamefont{N.~H.}~\bibnamefont{Barton}},
  \bibinfo{journal}{Genetics} \textbf{\bibinfo{volume}{140}},
  \bibinfo{pages}{821} (\bibinfo{year}{1995}).

\bibitem[{\citenamefont{\c{C}inlar}(1975)}]{cinlar-1975}
\bibinfo{author}{\bibfnamefont{E.}~\bibnamefont{\c{C}inlar}},
  \emph{\bibinfo{title}{Introduction to Stochastic Processes}}
  (\bibinfo{publisher}{Prentice Hall}, \bibinfo{address}{Englewood Cliffs},
  \bibinfo{year}{1975}), \bibinfo{edition}{1st} ed.

\bibitem[{\citenamefont{Johnson and Barton}(2002)}]{johnson-barton-2002}
\bibinfo{author}{\bibfnamefont{T.} \bibnamefont{Johnson}} \bibnamefont{and}
  \bibinfo{author}{\bibfnamefont{N.~H.} \bibnamefont{Barton}},
  \bibinfo{journal}{Genetics} \textbf{\bibinfo{volume}{162}},
  \bibinfo{pages}{395} (\bibinfo{year}{2002}).

\bibitem[{\citenamefont{Maia et~al.}(2003)\citenamefont{Maia, Botelho,
  and Fontanari}}]{maia-et-al-2003}
\bibinfo{author}{\bibfnamefont{L.~P.} \bibnamefont{Maia}},
  \bibinfo{author}{\bibfnamefont{D.~F.}~\bibnamefont{Botelho}}, \bibnamefont{and}
  \bibinfo{author}{\bibfnamefont{J.~F.} \bibnamefont{Fontanari}},
  \bibinfo{journal}{J. Math. Biol.} \textbf{\bibinfo{volume}{47}},
  \bibinfo{pages}{453} (\bibinfo{year}{2003}).

\bibitem[{\citenamefont{Maia}(2005)}]{maia-2005}
\bibinfo{author}{\bibfnamefont{L.~P.}~\bibnamefont{Maia}},
  \bibinfo{journal}{J. Math. Biol.} \textbf{\bibinfo{volume}{51}},
  \bibinfo{pages}{114} (\bibinfo{year}{2005}).

\bibitem[{\citenamefont{Chao}(1990)}]{chao-1990}
\bibinfo{author}{\bibfnamefont{L.}~\bibnamefont{Chao}},
  \bibinfo{journal}{Nature} \textbf{\bibinfo{volume}{348}},
  \bibinfo{pages}{454} (\bibinfo{year}{1990}).

\end{thebibliography}
\end{document}